\documentclass[a4paper,12pt]{article}

\usepackage{graphicx}
\usepackage{amsmath}

\begin{document}
\noindent {\bf \large Velocity, acceleration and gravity in Einstein's relativity}
\vskip1.0truecm
\noindent {\bf Marek A. Abramowicz}
\vskip0.2truecm \noindent
{\it Physics Department, G{\"o}teborg University, Sweden \,\&\, N.Copernicus Astronomical Centre PAN, Warsaw, Poland,}
 {\tt marek.abramowicz@physics.gu.se}
\vskip0.6truecm
\noindent {\footnotesize {\bf Abstract:} Einstein's relativity theory demands that all meaningful physical objects should be defined covariantly, i.e. in a coordinate independent way. Concepts of relative velocity, acceleration, gravity acceleration and
gravity potential are fundamental in Newton's theory and they are imprinted in everyone's physical intuition. Unfortunately, relativistic definitions of them are not commonly known or appreciated. Every now and then some confused authors use wrong, non-covariant, definitions of velocity, acceleration and gravity, based on their vague Newtonian intuitions and hidden in a
superficial, often purely semantic, relativistic disguise. A recent example of such a confusion (Gorkavyi \& Vasilkov, 2016) is discussed at the end of this Note.}
\vskip1.0truecm
%
\noindent {\bf 1. Introduction}
%
\vskip0.1truecm
\noindent Einstein's general relativity is already a century old. It brought a profound insight into the very nature of physical reality, in particular of space, time and gravity. Numerous important issues concerning specific problems in physics, astrophysics and cosmology have been fully understood and explained in every mathematical detail. However, Einstein's relativity
still attracts mobs of naive amateurs who are convinced that ``Einstein was wrong'' and that they may demonstrate this (e.g. that time must be absolute) in a few lines of a simple calculation. More sophisticated ones, often professional physicists, know that
Einstein was right, but they claim that there are important overlooked phenomena in his theory (e.g. a repulsive gravity in the Schwarzschild space-time) which they have found --- again in a few lines of a simple calculation.
\vskip0.1truecm
\noindent These false claims are usually based on rejecting one or two of the fundamental principles of Einstein's general relativity --- the general covariance or the equivalence principle. General covariance demands that all meaningful physical objects should be defined covariantly, i.e. in a coordinate independent way. The equivalence principle says that acceleration of the reference frame cannot be locally distinguished from gravity. Thus, locally gravity is a fictitious phenomenon, it is caused by acceleration of the reference frame; the true gravity appears only non-locally and is connected to the Riemannian  curvature of space-time.

\vskip3.0truecm
%
\noindent {\bf 1. The relative velocity}
%
\vskip0.1truecm
\noindent In Einstein's theory, the four-velocity $u^i$ of a particle is an absolute space-time object. Let $n^i$ denote the four-velocity of an observer (again, an absolute space-time object). The proper covariant way to define a velocity $v^i$ relative to the observer $n^i$ is by projection $h^i_{~j}$ of $u^i$ into the local rest frame of the observer,
\begin{equation}
v^i = h^i_{~j} u^j; ~~{\rm where}~~ h^i_{~j} = \delta^i_{~j} - n^i n_j .
\label{three-velocity}
\end{equation}
Let us also  define two scalars $\gamma$ and $v$ by covariant definitions,
\begin{equation}
\gamma = u^i n_i, ~~ v^2 = - \frac{v^i v_i}{\gamma^2}.
\end{equation}
It is elementary to derive from these definitions the ``Lorentz'' form of the four-velocity,
\begin{equation}
u^i = \gamma \left( n^i + v \tau^i \right).
\label{four-velocity-Lorentz}
\end{equation}
Here $\tau^i$ is a unit vector (space-like) in the local rest frame of the observer $n^i$. One may say that $v$ is the ``speed'' and $\tau^i$ the ``direction'' of the velocity relative to the observer $n^i$. It is also elementary to check that
\begin{equation}
\gamma = \frac{1}{\sqrt{1 - v^2}},
\end{equation}
i.e. that $\gamma$ and $v$ are the ``Lorentz'' quantities, familiar from Einstein's special relativity. To see how this invariant Lorentz form is useful in practical calculations, let us consider another observer $N^i$ with respect to whom the observer $n^i$ moves with a relative velocity $V$, i.e.
\begin{equation}
n^i = \Gamma \left( N^i + V T^i \right), ~~ \Gamma = \frac{1}{\sqrt{1 - V^2}}.
\label{another-observer}
\end{equation}
Here $T^i$ is a unit time-like vector. Inserting (\ref{another-observer}) into (\ref{four-velocity-Lorentz}) together with $\tau^i$ expressed by $N^i$ and $T^i$, one gets after a short and elementary calculation\footnote{Hint: assuming that all motions considered here are collinear, one has $\tau^i = \alpha N^i + \beta T^i$. One may get two unknown coefficients $\alpha$ and $\beta$ by solving two equations $\tau^i \tau_i = -1$, and $\tau^i n_i =0$, but it is easier to guess the solution: $\tau^i = \Gamma(V N^i + T^i)$. },
\begin{equation}
u^i = {\tilde \gamma} \left( N^i + {\tilde v} T^i \right), ~~{\tilde v} = \frac{v + V}{1 + v\,V}, ~~{\tilde \gamma} =
\frac{1}{\sqrt{1 - {\tilde v}^2}},
\label{Lorentz-addition}
\end{equation}
i.e. the Lorentz rule of adding velocities. Although most of the special relativity kinematics may be explained to students in this simple, invariant geometric way (gently preparing them for challenges of general relativity), most lecturers and textbooks prefer to use an approach based on coordinates. One of a few exceptions is a remarkable and beautiful book by Gourgoulhon (2013).
\vskip0.3truecm
%
\noindent {\bf 2. The acceleration}
%
\vskip0.1truecm
\noindent Acceleration is invariantly defined as $a^i = u^k \nabla_k u^i$.
\vskip0.1truecm
\noindent Using the Lorentz form for the four-velocity (\ref{four-velocity-Lorentz}) one may write
\begin{equation}
a^i = u^k \nabla_k u^i = \gamma(n^k + v\tau^k) \nabla_k [\gamma(n^i + v\tau^i)].
\label{acceleration-definition}
\end{equation}
After a straightforward manipulation one may expand (\ref{acceleration-definition}) into,
\begin{eqnarray}
a^i = &&\gamma^2 (n^k \nabla_k n^i) \label{first} \\
    + &&\gamma^2 v (n^k \nabla_k \tau^i + \tau^k \nabla_k n^i) \label{second} \\
    + &&\gamma^2 v^2 (\tau^k \nabla_k \tau^i)  \label{third} \\
    + &&{\dot{\gamma}}n^i + {\dot{(\gamma v)}}\tau^i, \label{fourth}
\end{eqnarray}
where dot is defined as ${\dot X} = u^k \nabla_k X$. It is convenient to use the identity $\gamma^2 = 1 + \gamma^2 v^2$ in order to rewrite (\ref{first}), and then combine two terms containing $\gamma^2 v^2$. Finally, after projecting onto the local rest-space of the observer $n^i$, the acceleration formula takes the form
\begin{eqnarray}
a^i_{\perp} = &&n^k \nabla_k n^i \label{first-a} \\
    + &&\gamma^2 v (n^k \nabla_k \tau^i + \tau^k \nabla_k n^i)_{\perp} \label{second-a} \\
    + &&\gamma^2 v^2 (\tau^k \nabla_k \tau^i + n^k \nabla_k n^i)_{\perp}  \label{third-a} \\
    + &&\gamma^3{\dot{v}}\tau^i. \label{fourth-a}
\end{eqnarray}
Here the symbol $\perp$ denotes projection, $X^i_{\perp} = (\delta^i_{~k} - n^i n_k)X^k$. Obviously, $n^i_{\perp} = 0$, $\tau^i_{\perp} = \tau^i$ and
\begin{equation}
(n^k \nabla_k n^i)_{\perp} = n^k \nabla_k n^i \equiv g^i.
\label{perp}
\end{equation}
Note, that each of the four terms (\ref{first-a})-(\ref{fourth-a}) in the acceleration formula is covariantly defined. In addition, each term (\ref{first-a})-(\ref{fourth-a}) is defined {\it uniquely} -- for a specific observer $n^i$ and a specific four-velocity $u^i$ there is only one way to write the decomposition (\ref{first-a})-(\ref{fourth-a}) in terms of powers of $v$
and its derivative. This suggests that each of terms (\ref{first-a})-(\ref{fourth-a}) should have a well-defined physical meaning. This is indeed the case (Abramowicz, Nurowski \& Wex, 1993).

\vfill \eject \null
%
\noindent {\bf 3. Gravity}
%
\vskip0.1truecm
\noindent The absolute, covariant quantity $g^i$ that appears in (\ref{first-a}) and (\ref{perp}) is observer's $n^i$ acceleration, i.e. the acceleration of the reference frame. According to Einstein's {\it equivalence principle}, acceleration of the reference frame cannot be locally distinguished from gravity. Thus, in general, $g^i$ should be considered to be ``the gravity'', i.e. the ``gravitational acceleration''. In Newton's theory the gravitational acceleration equals (minus) the gradient of the gravitational potential $\Phi_N$,
\begin{equation}
{\rm Newtonian~formula\hskip-0.1truecm:}~~ g_i = - \nabla_i \Phi_N
\label{Newton-gravity}
\end{equation}
Stationary space-times admit a time-like Killing vector $\eta^i$,
\begin{equation}
\nabla_i \eta_k + \nabla_k \eta_i = 0 ~~({\rm Killing~equation}) \Rightarrow \eta^i \nabla_i \eta_k = -\frac{1}{2}\nabla_k \ln
(\eta^i \eta_i).
\label{Killing-equation}
\end{equation}
One may define the stationary observer $n^i$ by an invariant condition,
\begin{equation}
n^i = e^{-\Phi} \eta^i, ~~{\rm where}~~ \Phi = \frac{1}{2}\ln (\eta^i \eta_i) .
\label{static-observer}
\end{equation}
From the Killing equation (\ref{Killing-equation}) it then follow in one line of algebra,
\begin{equation}
g_i = n^k \nabla_k n_i = -\frac{1}{2}\nabla_i\ln(\eta^k \eta_k) = -\nabla_i\Phi.
\label{gravitational-acceleration}
\end{equation}
Comparing this with Newton's formula (\ref{Newton-gravity}), one may call $\Phi$ the relativistic gravitational potential\footnote{Of course, this potential is well-known. For example, Landau \& Lifshitz (1973) in equation (3) in exercise 1, in \textsection 88, noticed that $(a_i)_{\perp} = -\nabla_i\Phi$ for $v = 0$.}. The Killing vector $\eta^i$ defines time coordinate $t$ by $\eta^i = \delta^i_{~t}$. It is, obviously, $\eta^i\eta^k g_{ik} = g_{tt}$
and therefore the gravitational potential,
\begin{equation}
\Phi = \frac{1}{2}\ln(\eta^i\eta_i) = \frac{1}{2}\ln g_{tt}
\label{gravitational-potential}
\end{equation}
In the case of a weak gravitational field one has (see e.g. Landau \& Lifshitz, 1973; \textsection 87) $g_{tt} = 1 + 2\Phi_N$, where $\Phi_N \ll 1$ is the Newtonian gravitational potential. Therefore, in the weak field
\begin{equation}
\Phi = \Phi_N
\label{gravitational-potential-Newtonian-limit}
\end{equation}
as it should be.
\noindent Let us now consider a case of a space-time which is static and axially symmetric, i.e. that it posses two orthogonal and commuting Killing vectors $\eta^i$ and $\xi^i$. These vectors obey,
\begin{eqnarray}
&\nabla_i \eta_k + \nabla_k \eta_i = 0 = \nabla_i \xi_k + \nabla_k \xi_i,& \label{two-Killing} \\
&\eta^i \nabla_i \xi_k - \xi^i \nabla_i \eta_k = 0,&  \label{two-Killing-commute} \\
&\eta^i \xi_i = 0.& \label{two-Killing-orthogonal}
\end{eqnarray}
Let us define the specific energy $E$ and the specific angular momentum $\ell$ by invariant equations,
\begin{equation}
E = \eta^i u_i ~~~ \ell = -\frac{1}{E}\, \xi^i u_i
\label{specific-energy-momentum}
\end{equation}
From Killing equations (\ref{two-Killing}) it follows that $E$ and $\ell$ are constants of geodesic motion.
\vskip0.1truecm
\noindent The Schwarzschild space-time discussed in the next Section posses such vectors. In the Schwarzschild coordinates one has,
\begin{eqnarray}
&\eta^i = \delta^i_{~t}, ~~ \xi^i = \delta^i_{~\phi},& \nonumber \\
&\eta^i \eta_i = g_{tt} =(g^{tt})^{-1}, ~~ \xi^i \xi_i = g_{\phi\phi} = (g^{\phi\phi})^{-1},& \nonumber \\
&E = u_t, ~~ \ell = -\frac{u_\phi}{u_t}.&
\label{two-Killing-coordinates}
\end{eqnarray}
Here we do not assume any specific form of the metric, however. Let us consider a circular motion with constant orbital velocity
${\rm v}$,
\begin{equation}
u^i = \gamma(n^i + {\rm v} \tau^i), ~~~ \tau^i = \frac{1}{\sqrt{-\xi^k\xi_k}}\xi^i, ~~~{\dot {\rm v}}=0.
\label{circular-motion}
\end{equation}
The specific angular momentum $\ell$ and the orbital velocity ${\rm v}$ are linked by the relation
\begin{equation}
\ell = {\rm v}{\tilde r}, ~~~ {\tilde r} = \left( -\frac{\xi^i \xi_i}{\eta^k \eta_k}\right)^{1/2} = {\rm radius~of~gyration},
\label{velocity-momentum-relation}
\end{equation}
which has the same form as in Newton's theory. In Newton's theory the radius of gyration for an orbiting particle equals the distance from the axis of rotation. Using equations (\ref{two-Killing})-(\ref{circular-motion}) one brings the acceleration formula (\ref{first-a})-(\ref{first-a}) into the form\footnote{In the Kerr space-time the Killing vectors $\eta^i$ and $\xi^i$ commute, but they are not orthogonal. In such space-times the term (\ref{second-a}) is nonzero, it represents a Coriolis-Lense-Thirring acceleration.} ,
\begin{equation}
a_i = (a_i)_{\perp} = - \nabla_i \Phi + \frac{1}{\gamma^2}\left( \frac{{\rm v}^2}{\tilde r}\right)\nabla_i {\tilde r}.
\label{acceleration-circular}
\end{equation}
Calculations are straightforward, but one should be careful with signs. Formula (\ref{acceleration-circular}) has the same form
as the corresponding Newtonian formula, except that in Newton's theory one has $\gamma = 1$. All quantities that appear in
formula (\ref{acceleration-circular}) are covariantly defined and have the same physical meaning as their Newtonian counterparts.
There is, however an important difference here. In Newton's theory, the radial gradients of $-\Phi = GM/r$ and ${\tilde r} = r$
have always (i.e. for each $r > 0$) opposite signs, and therefore there always exist circular geodesics $a_i = 0$ (in Newton's
theory called Keplerian orbits). In Einstein's theory it is not so --- for example, in the Schwarzschild space-time, the two
gradients have opposite signs for $r > 3M$ but the same signs for $r < 3M$. {\it Thus, there are no circular geodesics for} $r <
3M$. At $r = 3M$, where the geodesic circular photon trajectory is located $\nabla_i{\tilde r} = 0$ and therefore the non-zero
acceleration for non-geodesic particles moving along $r = 3M$ does not depend on the orbital speed. This is a real physical
effect discovered by Abramowicz \& Lasota (1974) and formally explained in terms of the optical geometry by Abramowicz, Carter
and Lasota (1988). The effect, known as the ``centrifugal force reversal'' correctly and surely guides one's intuition into
understanding of several non-Newtonian effect in strong gravity. For its popular explanation see a {\it Scientific American}
article by Abramowicz (1974).
\vskip0.1truecm
\noindent In Newton's classical mechanics one introduces a useful concept of the ``effective potential'' ${\cal U}$. Acceleration
is equal minus a partial derivative of ${\cal U}$ at constant angular momentum $\ell$,
\begin{equation}
a_r = - \left( \frac{\partial {\cal U}}{\partial r} \right)_{\ell = {\rm const}}.
\label{acceleration-effective-potential}
\end{equation}
Integration of (\ref{acceleration-circular}) shows that in our case one should define the relativistic effective potential
by\footnote{Hint: it is easier to make a direct check by differentiating ${\cal U}$ in (\ref{acceleration-effective-potential})
to get -RHS of equation (\ref{acceleration-circular}).}
\begin{equation}
{\cal U} = \Phi - \frac{1}{2} \ln \left( 1 -\frac{\ell^2}{{\tilde r}^2}  \right).
\label{relativistic-effective-potential}
\end{equation}
One calculates Newtonian limit ${\cal U}_N$ of (\ref {relativistic-effective-potential}) by using
(\ref{gravitational-potential-Newtonian-limit}) and expanding with respect to $\ell^2/{\tilde r}^2 = {\rm v}^2 \ll 1$,
\begin{equation}
{\cal U} = \Phi - \frac{1}{2} \ln \left( 1 -{\rm v}^2  \right) = \Phi_N + \frac{1}{2}{\rm v}^2.
\label{effective-potential-Newtonian-limit}
\end{equation}
Let me show another, non-trivial, example of the physical correctness, and a remarkable practical usefulness, of the covariant
definitions of reference frame, velocity, gravitational acceleration, gravitational potential and effective potential which we
have adopted here. Let us consider a geodesic, but slightly non-circular, orbital motion. In the Schwarzschild space-time such
orbits are planar, i.e. with $E = u_t \not = 0$, $-E\ell =u_\phi \not = 0$, $u^r = {\dot r} \not = 0$ and $u^\theta = 0$. The
almost circular orbits may be expressed in terms of a fixed radius $r_*$ and its varying in time $s$ proper time ``perturbation''
$\delta r(s) \ll r_*$,
\begin{equation}
r = r_* + \delta r(s) \Rightarrow {\dot r} = \dot{\delta r}.
\label{circular-orbit-radius}
\end{equation}
For such orbits we may write step by step, (\ref{circular-orbit-calculation-1}) $\Rightarrow$
(\ref{circular-orbit-calculation-2}) $\Rightarrow$ (\ref{circular-orbit-calculation-3}) $\Rightarrow$
(\ref{circular-orbit-calculation-4}),
\begin{eqnarray}
&1 = u^i u^k g_{ik} = u_i u_k g^{ik} = E^2 (g^{tt} + \ell^2 g^{\phi \phi}) + {\dot{\delta r}}^2 g_{rr}&
\label{circular-orbit-calculation-1} \\
& 1 - g_{rr}{\dot{\delta r}}^2 = E^2 (\eta^i\eta_i)^{-1}\left( 1 - \frac{\ell^2}{{\tilde r}^2}\right)&
\label{circular-orbit-calculation-2}\\
&\frac{1}{2}\ln (1 - g_{rr}{\dot{\delta r}}^2) = {\cal E} - {\cal U}, ~~~ {\rm where} ~~~ {\cal E} = \ln E &
\label{circular-orbit-calculation-3} \\
&-g_{rr}\,\frac{1}{2}\,{\dot{\delta r}}^2 = {\cal E} - {\cal U} \label{circular-orbit-calculation-4}&
\end{eqnarray}
Note how logarithms are important here! The step
(\ref{circular-orbit-calculation-3})$\Rightarrow$(\ref{circular-orbit-calculation-4}) was made by expanding the logarithm on the
LHS of (\ref{circular-orbit-calculation-3}) and taking into account that  ${\dot{\delta r}} \ll 1$. Imagine a perturbation of an
initially strictly circular orbit by adding to its energy ${\cal E}$ a small bit $\delta {\cal E}$ and keeping the angular
momentum $\ell$ constant. Note that because both initial and perturbed orbits are geodesic, it is $\dot{\delta{\cal E}} = 0$.
\begin{eqnarray}
&-g_{rr}\frac{1}{2}{\dot{\delta r}}^2 = \delta {\cal E} - \frac{1}{2}\left( \frac{\partial^2 {\cal U}}{\partial r^2} \right)_\ell
(\delta r)^2.&  \label{perturbation-1}\\
& (\ddot{\delta r}) + \omega^2 (\delta r) = 0.& \label{perturbation-2}\\
&\omega^2 = \left( \frac{\partial^2 {\cal U}}{\partial R^2} \right)_\ell. & \label{perturbation-3}
\end{eqnarray}
Here $R$ is the proper length in the radial direction, defined (invariantly) by $dR^2 = g_{rr} dr^2$. In (\ref{perturbation-1}), in the Taylor expansion of $\delta{\cal U}$, the first derivative term was dropped because $(\partial_r {\cal U})_{\ell} = 0$ for the unperturbed circular geodesic. The step (\ref{perturbation-1})$\Rightarrow$(\ref{perturbation-2}) was made by differentiation of (\ref{perturbation-1}) by $d/ds$ side by side. The resulting harmonic oscillator equation (\ref{perturbation-2}) has the same form as in Newton's theory. The epicyclic frequency formula (\ref{perturbation-3}) also has the Newton form. 
\vskip0.1truecm
\noindent It is important to realize that the epicyclic frequency $\omega$ is invariantly connected to the {\it true} gravity i.e. to the Riemannian curvature of the space-time. To see this, let us  consider the geodesic deviation equation,
\begin{equation}
\frac{d^2 \epsilon^i}{ds^2} = R^i_{~kjn}u^k u^j \epsilon^n
\label{geodesic-deviation}
\end{equation}
where $R^i_{~kjn}$ is the Riemann curvature tensor, $u^k = u^k(s, \mu)$ is a family of circular geodesic and $\epsilon^i = (\delta \mu)(du^i/d\mu)$ is an infinitesimal vector connecting two infinitesimally close geodesics. In the Schwarzschild space-time this geodesic deviation equation has the specific form, 
\begin{equation}
\ddot{(\delta \mu)} + \omega^2 (\delta \mu) = 0, 
\label{geodesic-deviation-Schwarzschild}
\end{equation}
where the epicyclic frequency determined by the geodesic deviation equation,
\begin{equation}
\omega^2 = (u^t)^2\left(R_{rtrt} +\frac{\ell^2}{{\tilde r}^4} \ R_{r\phi r \phi}\right) = \left( \frac{\partial^2 {\cal U}}{\partial R^2} \right)_\ell.
\label{geodesic-deviation-Schwarzschild-01}
\end{equation}
is the same, as the epicyclic frequency previously calculated as the the second derivative of the effective potential ${\cal U}$. Thus, the effective potential ${\cal U}$, and therefore also the gravitational potential $\Phi$, are both uniquely linked to the {\it true gravity} given by the Riemann curvature of the space-time.
\vskip0.1truecm
\noindent The epicyclic frequency (\ref{perturbation-3}) refers to a co-moving observer. The epicyclic frequency $\omega_*$ measured by a distant astronomer equals,
\begin{equation}
\omega_* = e^\Phi\,\gamma^{-1}\, \omega
\label{epicyclic-requency-at-infinity}
\end{equation}
In both Newton's theory and Einstein's relativity in stationary, axially symmetric space-times, the orbital frequency $\Omega = d\phi/dt = \ell/{\tilde r}^2$. If $\Omega = \omega_*$, as it is in Newton's theory with the $\Phi_N = -GM/r$ gravitational potential, then orbits are {\it closed}, Keplerian ellipses. If $\Omega \not = \omega_*$, as it is in Einstein's theory, the
orbits are not closed --- the ellipses precess, with  the perihelion advance per one revolution given by,
\begin{equation}
\Delta \phi = 2\pi \frac{\Omega - \omega_*}{\Omega}
\label{perihelion-advance}
\end{equation}
\vskip0.3truecm
%
\noindent {\bf 5. A misconception about velocity, acceleration and gravity}
%
\vskip0.1truecm
\noindent Some authors are gravely confused about proper definitions of velocity and acceleration in Einstein's general
relativity. A recent paper by Gorkavyi \& Vasilkov (2016) is a particularly bizarre example of such a confusion. Among other
issues, the paper is reconsidering (wrongly) a classic and long understood problem of radial motion in the Schwarzschild metric,
\begin{equation}
ds^2 = \left( 1 - \frac{r_0}{r}\right)dt^2 - \left( 1 - \frac{r_0}{r}\right)^{-1} dr^2 - r^2 d\theta^2 - r^2\sin^2\theta\,
d\phi^2.
\label{Schwarzschild-metric}
\end{equation}
Here $r_0 = 2M$, and $M$ is the mass of a spherically symmetric source of the metric (\ref{Schwarzschild-metric}), e.g. a star or
a black hole. In this later case $r = r_0$ marks the location of the black hole event horizon. Here, we restrict attention to a
region $r > r_0$. A proper covariant definition of the radial speed $v$ in the stationary observer frame (\ref{static-observer})
is given by the ``Lorentz'' four-velocity formula discussed in Section 1,
\begin{equation}
u^i = \gamma (n^i + v \tau^i), ~~~n^i = \frac{1}{(\eta^i\eta_i)^{1/2}}\eta^i, ~~~\tau^i = \frac{1}{(-g_{rr})^{1/2}}\delta^i_{~r}.
\label{radial-velocity}
\end{equation}
Here $\eta^i$ is the time-like Killing vector. Note, that from the second equation in (\ref{radial-velocity}) and discussion in
the previous Section it follows that the gravitational potential is defined by
\begin{equation}
\Phi = \frac{1}{2}\ln\left( 1 - \frac{r_0}{r} \right).
\label{potential-Schwarzschild-ge-te-te}
\end{equation}
Note also, that $E = \eta^i u_i$ is the conserved ``energy'', i.e. a constant of geodesic motion. The proper covariant way to
define radial component of acceleration is by
\begin{equation}
a \equiv a^r = \delta^r_{~i} \left( u^k \nabla_k u^i\right).
\label{radial-acceleration}
\end{equation}
Gorkavyi \& Vasilkov (2016) define ``radial velocity'' $V$ and ``radial acceleration'' $\alpha$ by coordinate-dependent,
non-covariant, conditions,
\begin{equation}
V = \frac{dr}{dt}, ~~~ \alpha = \frac{d^2r}{dt^2}.
\label{velocity-acceleration-coordinate}
\end{equation}
This by itself would not be a problem. One may of course use whatever name for whatever quantity that appears in calculations, as
long as calculations are formally correct. The problem is that Gorkavyi \& Vasilkov (2016) treat the arbitrary names introduced
by themselves (velocity, acceleration, gravity) far too seriously. Their claim that there is a ``repulsive gravitational force''
in the Schwarzschild metric is only based on improper use of words, not on physics. When they apply this to cosmology and claim
that their repulsive gravity may explain the accelerated expansion of the Universe with no need for a dark energy, they again
play with abused meaning of words. Abramowicz \& Lasota (2016) proved, by appealing to the Friedmann equations, that the cosmic
claim made by Gorkavyi \& Vasilkov (2016) is totally groundless. Abramowicz \& Lasota (2016) have not discussed in detail some
grave mistakes made by Gorkavyi \& Vasilkov (2016) in the Schwarzschild metric context. This I do here.
\vskip0.1truecm
\noindent Let us multiply the first equation in (\ref{radial-velocity}) side by side by $\eta^i$
\begin{equation}
E = \gamma (\eta^i n_i) = \gamma \left( 1 - \frac{r_0}{r}\right)^{1/2}, ~~{\rm i.e.}~~ \gamma = \frac{E}{\left( 1 -
\frac{r_0}{r}\right)^{1/2}}.
\label{energy-gamma}
\end{equation}
This means that all particles, independently on their energy, approach the horizon with the speed of light, as measured by the
static observer, which is of course reasonable and understandable,
\begin{equation}
\gamma \rightarrow \infty, ~~~ {\rm and} ~~~ v \rightarrow 1 ~~{\rm for}~~  r \rightarrow r_0
\label{approach-horizon}
\end{equation}
From (\ref{radial-velocity}) and (\ref{Schwarzschild-metric}) one gets, after a few lines of algebra an expression for the radial
speed $v$ measured by the static observer,
\begin{equation}
v = \frac{(u^r)(-g_{rr})^{1/2}}{(u^t)(g_{tt})^{1/2}} = V \left(1 - \frac{r_0}{r} \right)^{-1},
\label{ve-and-deer-dete}
\end{equation}
From (\ref{approach-horizon}) one concludes that the ``velocity''
\begin{equation}
V \rightarrow 0 ~~{\rm for}~~  r \rightarrow r_0
\label{unphysical-unphysical}
\end{equation}
for particles approaching the horizon: this is an obviously unphysical behavior of an obviously unphysical quantity! However,
Gorkavyi \& Vasilkov (2016) seem to believe that this effect has something to do with, as they call it ``weakening of the
gravitational force'' or even ``repulsive gravitational force''. I admit that I do not follow their arguments based on semantics.
I~am rather surprised that they are treating these seriously, ignoring the clear and explicit statement of Zeldovich \& Novikov
(1967), which they know and even quote, that these effects are fictitious and depend on coordinate frame used.
\vskip0.1truecm
\noindent The authors not only use questionable, not covariant, definitions (\ref{velocity-acceleration-coordinate}) of velocity
and acceleration; in addition they calculate their ``acceleration'' for a geodesic motion in which, the true physical
acceleration is of course absent. They start from the coordinate form of the acceleration formula,
\begin{equation}
a^k = u^i \nabla_i u^k = u^i \partial_i u^k + \Gamma^k_{~ij}u^i u^j = \frac{d^2 x^k}{ds^2} +
\Gamma^k_{~ij}\frac{dx^i}{ds}\frac{dx^j}{ds}
\label{acceleration-coordinates}
\end{equation}
where the Christoffel symbol $\Gamma^k_{~ij}$ is defined by the coordinate derivatives of the metric,
\begin{equation}
\Gamma^k_{~ij} = \frac{1}{2} g^{mk}\left( \frac{\partial g_{im}}{\partial x^j} + \frac {\partial g_{jm}}{\partial x^i} -
\frac{\partial g_{ij}}{\partial x^m} \right).
\label{Christoffel}
\end{equation}
They assume geodesic motion, $a^k = 0$, and write (\ref{acceleration-coordinates}) in the form,
\begin{eqnarray}
\frac{d^2 r}{ds^2} = &&- \Gamma^r_{~ij}\frac{dx^i}{ds}\frac{dx^j}{ds}, \label{alpha-r}\\
\frac{d^2 t}{ds^2} = &&- \Gamma^t_{~ij}\frac{dx^i}{ds}\frac{dx^j}{ds}, \label{alpha-t}
\end{eqnarray}
to calculate their ``acceleration'' $\alpha$ for the radial geodesic motion. In addition to (\ref{alpha-r}) and (\ref{alpha-t}),
they use
\begin{equation}
\frac{dr}{ds} = \frac{dt}{ds} \frac{dr}{dt} = u^t \frac{dr}{dt}
\label{de-es-de-te}
\end{equation}
I have calculated $\alpha$ for a general radial motion, with no assumption that $a^i = 0$. Calculations are straightforward and yield
\begin{equation}
\alpha = a\,X - V^2 \left[ \Gamma^r_{rr} - 2
\Gamma^t_{tr}\right] - \Gamma^r_{tt}.
\label{alpha-acceleration}
\end{equation}
where the exact form of $X$ is not important here. Gorkavyi \& Vasilkov (2016) derived the same formula; the only difference is that in their formula the first term, $a\,X$, on the RHS of
(\ref{alpha-acceleration}) is absent because they assume $a=0$. 
%
Gorkavyi \& Vasilkov (2016) claim that the last term on the RHS of (\ref{alpha-acceleration})
\begin{equation}
- \Gamma^r_{tt} = \frac{1}{2g_{rr}}\frac{\partial g_{tt}}{\partial r}
= -\frac{1}{2}\left(1 - \frac{r_0}{r}\right) \frac{\partial}{\partial r}\left(1 - \frac{r_0}{r}\right)
\label{Gorkavyi-gravity}
\end{equation}
represents the gravitational acceleration. Therefore according to them, the gravitational acceleration formula is
\begin{equation}
\boxed{g = \frac{1}{2g_{rr}}\frac{\partial g_{tt}}{\partial r}}~~ {\rm
instead~of~the~correct~formula}~(\ref{gravitational-acceleration}), ~~g = -\frac{1}{2g_{tt}}\frac{\partial g_{tt}}{\partial r}.
\label{Gorkavyi-acceleration}
\end{equation}
Abramowicz \& Lasota (2016) called politely this mistake ($g_{rr}$ instead of the correct $g_{tt}$) ``a printing error''.
However, it seems that it was not just a printing error --- Gorkavyi \& Vasilkov (2016) are genuinely confused about the issue
here\footnote{At {\it http://don-beaver.livejournal.com/175194.html} they defend their choice of the formula for the
gravitational acceleration (in Russian).}.

Below I write the Gorkavyi \& Vasilkov (2016) ``gravitational potential'' $\Psi$ together with the correct expression for the
physically meaningful potential $\Phi$, discussed at length previously,
\begin{eqnarray}
{\rm Gorkavyi~and~Vasilkov~potential}\hskip-0.1truecm:~~  \Psi &=& \frac{1}{4}\left(1 - \frac{r_0}{r}\right)^2,
\label{Gorkavyi-potential} \\
{\rm Static~observer~potential}\hskip-0.1truecm:~~ \Phi &=& \frac{1}{2} \ln \left(1 - \frac{r_0}{r}\right) \label{static-potential}
\end{eqnarray}
The standard, covariantly defined, static observer gravitational potential describes correctly, according to its name, all
aspects of relativistic gravity, as we have shown in the previous Section. Its name and physical meaning follows directly from
Einstein's equivalence principle. The Gorkavyi and Vasilkov ``potential'' is not covariantly defined and it contradicts the
equivalence principle.
\vskip0.3truecm
%
\noindent {\bf 6. Conclusions}
%
\vskip0.1truecm
\noindent The paper by Gorkavyi \& Vasilkov (2016) is not even wrong, it is simply a nonsense. The only question that may be of
interest here is --- {\it how such a paper could possibly slip into the Monthly Notices of the Royal astronomical Society}? I may
offer just one explanation. I was told a story, perhaps apocryphal, about a famous referee report (many years ago) by Jim
Pringle: ``The paper is so obviously wrong that it should be published as it stands".

\vskip0.3truecm
\vfill \eject \null
\noindent {\bf References}                                                 {\small
\vskip0.2truecm \noindent
Abramowicz M.A., 1974, Scientific American, March volume
\vskip0.1truecm \noindent
Abramowicz M.A. \& Lasota J.-P., 1974, Acta Phys. Polon., B5, 327
\vskip0.1truecm \noindent
Abramowicz M.A., Carter B. \& Lasota J.-P., 1988, Gen. Relat. Grav., 20, 1173
\vskip0.1truecm \noindent
Abramowicz M.A. \& Lasota J.-P., 2016, http://arxiv.org/pdf/1608.02882v1.pdf
\vskip0.1truecm \noindent
Abramowicz M.A., Nurowski P. \& Wex N., 1993, Class. Quantum Grav., 10, L183
\vskip0.1truecm \noindent
Gorkavyi N. \& Vasilkov N., 2016, MNRAS {\bf 461}, 2929-2933
\vskip0.1truecm \noindent
Gourgoulhon E., 2003, {\it Special Relativity in General Frames}, Springer
\vskip0.1truecm \noindent
Landau L.D. \& Lifshitz E.M., 1973, {\it Field Theory} 6th ed., Nauka, Moscow
\vskip0.1truecm \noindent
Zeldovich Ya. B., Novikov I. D., 1967, {\it Relativistic Astrophysics}, Nauka, Moscow.
}    
\vskip1truecm \hrule \hrule \vskip0.4truecm
\noindent {\footnotesize This work was supported by the Polish National Research Center (NCN) grant No. 2015/19/B/ST9/01099.}

\end{document}